\documentclass[twoside]{elsart}
\usepackage{epsfig}
\usepackage{array}
\usepackage{amssymb,amsbsy,amsmath}
\def\bra#1{\langle \, {#1} \, | \;}
\def\ket#1{\; | \, {#1} \, \rangle}
\newcommand{\braket}[2]{\langle \, {#1} \, | \, {#2} \, \rangle}
\newcommand{\op}[1]{%
    \fontdimen12\textfont3=2pt\fontdimen12\scriptfont3=1.4pt%
    \!\null\mathop{\vphantom{#1}\smash{#1}}\limits_{\sim}\null\!}
\newcommand{\dint}{\mbox{d}}

\newcommand{\petaeins}{p_{\eta_1}}
\newcommand{\petazwei}{p_{\eta_2}}
\newcommand{\Mean}[1]{\big\langle\big\langle \; {#1}\; 
            \big\rangle\big\rangle}
\newcommand{\SmallMean}[1]{\langle\langle {#1} 
            \rangle\rangle}

\newcommand{\EinsOp}
           {\;\smash{\raisebox{-1.1ex}{$\!\!\stackrel{\!\mbox{1}
            \hspace{-0.4ex}\rule[0.0ex]{0.06ex}{1.60ex}}{\sim}$}}}
\newcommand{\ddt}{\frac{d}{dt}}

\begin{document}
%---------- Titel und Abstract-------------------------------
%
\typeout{ --- >>> Nos\'e thermostat 2 particles <<<  --- }
\typeout{ --- >>> Nos\'e thermostat 2 particles <<<  --- }
\typeout{ --- >>> Nos\'e thermostat 2 particles <<<  --- }
%
%---------- Journal-----------------------------------------
%
\journal{PHYSICA A}
%\volume{}
%\issue{}
%\pubyear{}
%\firstpage{}
%\lastpage{}
\begin{frontmatter}
\title{Nos\'e-Hoover sampling of quantum entangled distribution functions}
\author{D. Mentrup \and J. Schnack\thanksref{JS}}
\address{Universit\"at Osnabr\"uck, Fachbereich Physik \\ 
         Barbarastr. 7, D-49069 Osnabr\"uck}

\thanks[JS]{email: jschnack\char'100uos.de,\\
            WWW:~http://www.physik.uni-osnabrueck.de/makrosysteme}

\begin{abstract}

\noindent
While thermostated time evolutions stand on firm grounds and are
widely used in classical molecular dynamics (MD) simulations
\cite{Tuck00}, similar methods for \emph{quantum} MD schemes are
still lacking. In the special case of a quantum particle in a
harmonic potential, it has been shown that the framework of
coherent states permits to set up equations of motion for an
isothermal quantum dynamics \cite{Men01}.  In the present
article, these results are generalized to indistinguishable
quantum particles. We investigate the consequences of the
(anti-)symmetry of the many-particle wavefunction which leads to
quantum entangled distribution functions. The resulting
isothermal equations of motion for bosons and fermions contain
new terms which cause Bose-attraction and
Pauli-blocking. Questions of ergodicity are discussed for
different coupling schemes.

\vspace{1ex}
\vspace{1ex}

\end{abstract}
\end{frontmatter}
\raggedbottom
%%%%%%%%%%%%%%%%%%%%%%%%%%%%%%%%%%%%%%%%%%%%%%%%%%%%%%%%%%%%%%%%%%%%%%%%
\section{Introduction and summary}

Classical MD simulations are performed by solving Hamilton's equations
of motion numerically. Therefore, the internal energy of the system is
conserved during time evolution. If the ergodic hypothesis is
satisfied, a time average of a macroscopic observable is, under
equilibrium conditions, the same as a microcanonical ensemble average.
Hence, isoenergetic molecular dynamics entails a method for the
calculation of microcanonical ensemble averages.

In the canonical ensemble, however, the internal energy of the system
is not constant, but free to fluctuate by thermal contact to an
external heat bath. In order to adapt classical MD simulations to the
problem of the calculation of canonical ensemble averages, i.~e.~to
pass over from an isoenergetic to an isothermal time evolution,
powerful methods have been developed since the 1980s, and they are
commonly used nowadays \cite{Hoo85,Nos91}. Some of them, like the
Nos\'e-Hoover chain technique \cite{Mar92} or the
Kusnezov-Bulgac-Bauer thermostat \cite{KBB90}, are based on
deterministic time-reversal equations of motion leading to
trajectories that fill the phase space of the system according to the
canonical thermal weight. This allows the calculation of canonical
ensemble averages by means of molecular dynamics.

In quantum mechanics, the problem is more involved, and the pioneering
approaches of Grilli and Tosatti \cite{GrT89} and Kusnezov
\cite{Kus93} were not applied very much. Moreover, the quantum
mechanical time evolution itself is a hard computational problem.
However, a number of approximate quantum MD methods are available
\cite{RMP}. It is an open question whether they can be modified in a
manner appropriate to permit the calculation of quantum canonical
averages.

In a different methodological approach to isothermal quantum dynamics,
we have shown that in the special case of a quantum particle in a
harmonic oscillator potential, the framework of coherent states
permits to set up equations of motion for an isothermal quantum
dynamics \cite{Men01}. Following the approaches of Nos\'e and Hoover
or the similar KBB-technique, time-dependent pseudofriction terms are
added to the equations of motion for the parameters of coherent
states. The dynamics of the pseudofriction coefficients is designed
such that the desired thermal weight function is a stationary solution
of a generalized Liouville equation on a mixed quantum-classical phase
space.

The principle of indistiguishability of identical particles adds
an important quantum aspect to the problem of isothermal quantum
dynamics. The present article focusses on the modifications of
the method presented in \cite{Men01} for a system of two
non-interacting identical particles. The (anti-)symmetry of the
wavefunction leads to an entangled distribution function on
quantum phase space.  Surprisingly, we find that the originally
classical Nos\'e-Hoover approach succeeds to allow for this
quantum feature. The entanglement leads to additional terms in
the equations of motion of the pseudofricional coefficients that
act effectively like an attractive (for bosons) or repulsive
force (for fermions). We examine whether the modified equations
of motion lead to an ergodic time evolution, and it turns out
that the additional terms improve the overall ergodicity of the
various schemes. In addition, standard techniques to generate
ergodicity known from the classical methods are employed and
tested.  As a general result, ergodic time evolution is
achieved even for low temperatures without serious
difficulties. However, it is indispensable to thermalize both
particles since the effective forces do not suffice to yield an
ergodic time evolution of the whole system if only one particle
is coupled to a thermostat.
%%%%%%%%%%%%%%%%%%%%%%%%%%%%%%%%%%%%%%%%%%%%%%%%%%%%%%%%%%%%%%%%%%%%%%%%
\section{Method and setup}

\subsection{Two-particle distribution function in a harmonic
  oscillator potential}

The idea of the single-particle quantum Nos\'e-Hoover method
\cite{Men01} is to modify the equations of motion of the coherent
states \cite{Klauder} parameters $r$ and $p$ (which are combined to
the complex parameter $\alpha=\sqrt{\frac{m \omega}{2 \hbar}} \; r +
\frac{i}{\sqrt{2 m \hbar \omega}} \; p$) such that the quantum weight
function of a single particle in a harmonic oscillator potential,
denoted by $w_{qm}(r,p)$, is sampled in time. In precise analogy to
this case, we determine a thermal weight function
$w_\varepsilon^{(2)}(\alpha_1,\alpha_2)$ that permits to determine
canonical ensemble averages of two identical particles
\cite{PhDSchnack}. The values of $\varepsilon$, $+$ and $-$, refer to
bosons and fermions, respectively.

We write
\begin{align}
  \label{eq:1}
  \ket{A_\varepsilon}= 
  \left\{ 
    \begin{array}{l} \vspace{2mm} \op{S}_- \ket{\alpha_1,\alpha_2}
      \\
      \vspace{2mm} 
      \op{S}_+ \ket{\alpha_1,\alpha_2}
    \end{array} 
  \right. 
\end{align}
for the (anti-)symmetrized two-particle wavefunctions. $\op{S}_-$
denotes the antisymmetrizing projector (i.~e., $\ket{A_-}$ is a
two-particle Slater determinant), $\op{S}_+$ the symmetrizing
projector.

The starting point for the calculation of $w_\varepsilon^{(2)}$ is the
following expression for the calculation of a thermal expectation
value $\SmallMean{\op{B}}$ as a phase space integral,
\begin{align}
  \label{eq:2}
  \SmallMean{\op{B}} &=
  \frac{1}{Z^{(2)}_\varepsilon} \;
  \mbox{tr} \left( \op{B} e^{-\beta \op{H}} \right) 
  \\
  &=
  \frac{1}{Z^{(2)}_\varepsilon} 
  \iint \frac{\dint^2 \alpha_1}{\pi} \frac{\dint^2 \alpha_2}{\pi} \,
  \bra{\alpha_1, \alpha_2} 
  \op{S}^\dagger_\varepsilon \op{B} e^{-\beta \op{H}} 
  \ket{\alpha_1, \alpha_2}
  \nonumber
  \ ,
\end{align}
with $Z^{(2)}_\varepsilon=\mbox{tr}\left( e^{-\beta \op{H}} \right)$ being the
respective partition function. Note that we have dropped the projector
$\op{S}_\varepsilon$ acting upon the ket using the idempotency of
projectors. Decomposition of $e^{-\beta \op{H}}=e^{-\beta \op{H}/2} \,
e^{-\beta \op{H}/2}$ and a cyclic shift of the operators under the
trace enables further simplification, taking advantage of the specific
properties of coherent states \cite{Schnack99}. The following
expression
\begin{align}
  \label{eq:3}
  \SmallMean{\op{B}}
  &=
  \frac{1}{Z^{(2)}_\varepsilon} 
  \iint \frac{\dint^2 \alpha_1}{\pi} \frac{\dint^2 \alpha_2}{\pi} \,
  \underbrace{
    e^{-|\alpha_1|^2(e^{\beta\hbar\omega}-1)} 
    e^{-|\alpha_2|^2(e^{\beta\hbar\omega}-1)}
    \braket{A_\varepsilon}{A_\varepsilon}}
  _{\displaystyle w_\varepsilon^{(2)}(\alpha_1,\alpha_2)}
  \frac{\bra{A_\varepsilon}\op{B}\ket{A_\varepsilon}}
    {\braket{A_\varepsilon}{A_\varepsilon}}
\end{align}
finally defines $w_\varepsilon^{(2)}$ as the {\it thermal weight of
  the expectation value} $\frac{\bra{A_\varepsilon}\op{B}
  \ket{A_\varepsilon}}{\braket{A_\varepsilon}{A_\varepsilon}}$.

$w_\varepsilon^{(2)}(\alpha_1,\alpha_2)$ is not merely the product of
two one-particle thermal weight functions (as it would be the case for
two distinguishable particles), but contains in addition the factor
$\braket{A_\varepsilon}{A_\varepsilon}$ that accounts for the quantum
effects of indistinguishability. Moreover,
$w_\varepsilon^{(2)}(\alpha_1,\alpha_2)$ cannot be written as a
product of two functions depending only on $\alpha_1$ and $\alpha_2$,
respectively, since the term $\braket{A_\varepsilon}{A_\varepsilon}$
cannot be separated in this way. This is a result of the quantum
mechanical principle of indistinguishability. Therefore, we say that
$w_\varepsilon^{(2)}$ is \emph{entangled}.

In the case of fermions, we easily find
$\braket{A_-}{A_-}=\frac{1}{2}(1-e^{-|\alpha_1 - \alpha_2|^2})$. This
expression vanishes if $\alpha_1=\alpha_2$. A quantum state with two
identical fermions in the same one-particle-state is forbidden by the
Pauli exclusion principle, and therefore does not contribute to a
thermal average. In contrast, for bosons we have
$\braket{A_+}{A_+}=\frac{1}{2}(1+e^{-|\alpha_1 - \alpha_2|^2})$, which
contains the opposite sign that \emph{enhances} the thermal weight of
the quantum state with two bosons in the same one-particle state.

The thermal distribution function $w_\varepsilon^{(2)}$ yields the
correct partition function for the thermal average
$\SmallMean{\EinsOp}$. To show this, we calculate
\begin{align}
  \label{eq:4}
  \mbox{tr} \, e^{-\beta \op{H}} &=
  \iint \frac{\dint^2 \alpha_1}{\pi} \frac{\dint^2
    \alpha_2}{\pi}
  e^{-|\alpha_1|^2(e^{\beta\hbar\omega}-1)} 
  e^{-|\alpha_2|^2(e^{\beta\hbar\omega}-1)}
  \frac{1}{2} (1+ \varepsilon e^{|\alpha_1-\alpha_2|^2}) 
  \\
  &=
  \frac{1}{2}
  \left(
    \int \frac{\dint^2 \alpha_1}{\pi}
    e^{-|\alpha_1|^2(e^{\beta\hbar\omega}-1)}
  \right)^2 
  \nonumber 
  \\
  &\phantom{=}
  + \varepsilon \frac{1}{2}
  \left(
    \iint \frac{\dint^2 \alpha_1}{\pi} \frac{\dint^2
      \alpha_2}{\pi}
    e^{-|\alpha_1|^2(e^{\beta\hbar\omega}-1)} e^{-|\alpha_2|^2(e^{\beta\hbar\omega}-1)}
    \; e^{|\alpha_1-\alpha_2|^2}
  \right) 
  \nonumber 
  \\
  &=
  \frac{1}{2} 
  \left(
    \frac{1}{e^{\beta \omega}-1}
  \right)^2
  + \varepsilon \frac{1}{2}
  \left(
    \frac{1}{e^{2 \beta \omega}-1}
  \right) 
  \nonumber \\
  &=
  \frac{1}{2}
  \left[Z^{(1)}(\beta)^2 +\varepsilon Z^{(1)}(2 \beta) \right]
  \nonumber \ ,
\end{align}
which is a correct, well-known recursion relation for the two-particle
partition function. We indicate that the second integral is solved
most easily by a change of variables from $\alpha_1$, $\alpha_2$ to
$\alpha_+=\frac{1}{\sqrt{2}}(\alpha_1+\alpha_2)$,
$\alpha_-=\frac{1}{\sqrt{2}}(\alpha_1-\alpha_2)$, which leads to a
separation of the double integral.

\subsection{Modification of the equations of motion}

The Nos\'e-Hoover equations of motion for the coherent states
parameters that we will investigate read
\begin{align}
  \label{eq:7}
  \ddt r_1&= \frac{p_1}{m} \ , \qquad 
  \ddt r_2 = \frac{p_2}{m} \ ,
  \\
  \ddt p_1&= -m \omega^2 r_1 - p_1 \frac{\petaeins}{Q_1}
  \nonumber 
  \ ,
  \\
  \ddt p_2&= -m \omega^2 r_2 - p_2 \frac{\petazwei}{Q_2}
  \nonumber 
  \ .
\end{align}
The time dependence of the pseudofriction coefficients has to be
determined in a procedure analog to the case of a single particle,
i.~e.~we require that the desired distribution function
\begin{align}
  \label{eq:8}
  f_\varepsilon^{(2)}(\alpha_1,\alpha_2,\petaeins,\petazwei) 
  &\propto
  w_\varepsilon^{(2)}(\alpha_1,\alpha_2) 
  \exp \left( -\beta \Big(
    \frac{\petaeins^2}{2Q_1}+ \frac{\petazwei^2}{2Q_2} \Big)
    \right)
  \\
  &=
  e^{-(|\alpha_1|^2+|\alpha_2|^2)(e^{\beta\hbar\omega}-1)} 
  \frac{1}{2}(1 + \varepsilon e^{-|\alpha_1 - \alpha_2|^2})
\nonumber\\
&\qquad\times
  \exp \left( -\beta \Big(
    \frac{\petaeins^2}{2Q_1}+ \frac{\petazwei^2}{2Q_2} \Big)
    \right)
  \nonumber
  \\
  &=
  e^{-U} (1 \pm e^{-V})
  \nonumber
\end{align}
is a stationary solution of a generalized Liouville equation on the
phase space with elements $x=(\alpha_1,\alpha_2,\petaeins,\petazwei)
\equiv (r_1,p_1,r_2,p_2,\petaeins,\petazwei)$. The abbreviations $U$
and $V$ are defined as
\begin{align}
  \label{eq:9}
  U
  &=
  (|\alpha_1|^2+|\alpha_2|^2)(e^{\beta\hbar\omega}-1)
  +\beta \left( \frac{\petaeins^2}{2Q_1} 
    + \frac{\petazwei^2}{2Q_2} \right)
  \\
  &=
  \left( \frac{1}{2}m \omega^2 (r_1^2 + r_2^2)+\frac{1}{2m}(p_1^2+p_2^2)
  \right)
  \frac{e^{\beta\hbar\omega}-1}{\hbar\omega} 
  +\beta \left( \frac{\petaeins^2}{2Q_1} 
    + \frac{\petazwei^2}{2Q_2} \right)
  \ ,
  \nonumber \\[0.5cm]
  V
  &=
  |\alpha_1-\alpha_2|^2
  =
  \frac{m \omega}{2\hbar}(r_1-r_2)^2+\frac{1}{2m\hbar\omega}(p_1-p_2)^2
  \nonumber
  \ .
\end{align}
As in the one-particle case, $\petaeins$ and $\petazwei$ are regarded
as classical pseudofriction coefficients.

For the Liouville equation, we calculate
\begin{align}
  \label{eq:10}
  \ddt f_\varepsilon^{(2)}(\alpha_1,\alpha_2,\petaeins,\petazwei) =
  -(\dot{U}+\dot{V})f^\pm + \dot{V} e^{-U}
\end{align}
and
\begin{align}
  \label{eq:11}
  - \left(
    \frac{\partial}{\partial x} 
    \cdot \dot{x} \right)
  &=
  - \left(
    \frac{\partial}{\partial r_1} \dot{r}_1 +
    \frac{\partial}{\partial r_2} \dot{r}_2 +
    \frac{\partial}{\partial p_1} \dot{p}_1 + 
    \frac{\partial}{\partial p_2} \dot{p}_2 + 
    \frac{\partial}{\partial \petaeins} \dot{p}_{\eta_1} +
    \frac{\partial}{\partial \petazwei} \dot{p}_{\eta_2} 
    \right)
  \\
  &=
  \frac{\petaeins}{Q_1}+\frac{\petazwei}{Q_2}
  \nonumber
  \ ,
\end{align}
using the equations of motion \eqref{eq:7} and imposing, as common in
the present context \cite{KBB90}, the constraint $\partial \dot{p}_{\eta_i} /
\partial p_{\eta_i}=0$. 

After further transformations, we obtain the following equations of
motion for the pseudofriction coefficients from a comparison of the
coefficients of the terms $\petaeins/Q_1$ and $\petazwei/Q_2$ on both
sides of the Liouville equation:
\begin{align}
  \label{eq:12}
  \dot{p}_{\eta_1}
  &=\frac{1}{\beta} 
  \left( \frac{p_1^2}{m}
    \frac{e^{\beta\hbar\omega}-1}{\hbar\omega}-1 
    + \varepsilon p_1
    \frac{p_1-p_2}{m\hbar\omega}
    \frac{1}{e^V +\varepsilon 1}\right) 
  \ ,
  \\[1mm]
  \dot{p}_{\eta_2}
  &=
  \frac{1}{\beta}\left( \frac{p_2^2}{m}
    \frac{e^{\beta\hbar\omega}-1}{\hbar\omega}-1
    - \varepsilon p_2
    \frac{p_1-p_2}{m\hbar\omega}
    \frac{1}{e^V +\varepsilon 1}\right)
  \ .
  \nonumber
\end{align}
These equations, along with the set of equations \eqref{eq:7}, form a
genuine quantum Nos\'e-Hoover thermostat for two identical quantum
particles. The part $\frac{p_i^2}{m}
\frac{e^{\beta\hbar\omega}-1}{\hbar\omega}-1 \ , \; \, i=1,2 \, ,$ in
the equations of motion is familiar from the dynamics of a single
thermostatted particle. However, in the present case of two particles
we find additional terms that reflect the effects of Bose-attraction
and Pauli-blocking directly in the thermostatted dynamics, see next
section. The set of equations of motion \eqref{eq:7}, \eqref{eq:12}
conserves the quantity
\begin{align}
  \label{eq:13}
  H'=
  -\frac{1}{\beta} \ln w^{(2)}_{\varepsilon}
  (\alpha_1,\alpha_2) 
  +\frac{\petaeins^2}{2 Q_1}
  +\frac{\petazwei^2}{2 Q_2}
  +\frac{1}{\beta} \int^t \mbox{d}t' 
  \left( \frac{\petaeins(t')}{Q_1} + \frac{\petazwei(t')}{Q_2} 
  \right)
  \ .
\end{align}
The analogous approach using a KBB-scheme starts with the set of
equations
\begin{align}
  \label{eq:14}
  \frac{d}{dt}r_1 &= \frac{p_1}{m} -g_{r_1}'(\xi_1) F_{r_1}(r_1,p_1)
  \\
  \frac{d}{dt}p_1 &= -m \omega^2 r_1 - g_{p_1}'(\zeta_1)
  G_{p_1}(r_1,p_1)
  \nonumber 
  \\
  \frac{d}{dt}r_2 &= \frac{p_2}{m} -g_{r_2}'(\xi_2) F_{r_2}(r_2,p_2) 
  \nonumber 
  \\
  \frac{d}{dt}p_2 &= -m \omega^2 r_2 - g_{p_2}'(\zeta_2)
  G_{p_2}(r_2,p_2)
  \ ,
  \nonumber 
\end{align}
with the conserved quantity
\begin{multline}
  \label{eq:15}
  H'=
  -\frac{1}{\beta} \ln w_\varepsilon^{(2)}(\alpha_1,\alpha_2) 
  +\frac{g_{r_1}(\xi_1)}{\kappa_{r_1}}
  +\frac{g_{r_2}(\xi_2)}{\kappa_{r_2}}
  +\frac{g_{p_1}(\zeta_1)}{\kappa_{p_1}}
  +\frac{g_{p_2}(\zeta_2)}{\kappa_{p_2}} \\
  +\frac{1}{\beta} \int^t \mbox{d}t' 
  \left( 
    \frac{\partial F_{r_1}}{\partial r_1} g'_{r_1} +
    \frac{\partial F_{r_2}}{\partial r_2} g'_{r_2} +
    \frac{\partial G_{p_1}}{\partial p_1} g'_{p_1} +
    \frac{\partial G_{p_2}}{\partial p_2} g'_{p_2}
  \right)
  \ .
\end{multline}
The functions $g_{r_1}$, $g_{p_1}$, $g_{r_2}$, $g_{p_2}$ are chosen
such that the canonical distribution of the demons can be normalized.
The functions $F_{r_1}$, $F_{r_2}$, $G_{p_1}$, $G_{p_2}$ are
arbitrary. This scheme has the obvious advantage that positions and
momenta are treated symmetrically, i.~e.~pseudofriction coefficients
are present in all equations of motion. The time dependence of the
pseudofriction coefficients that is obtained from the Liouville
equation in the phase space with elements
$x=\{r_1,p_1,r_2,p_2,\xi_1,\xi_2,\zeta_1,\zeta_2\}$ reads
\begin{align}
  \label{eq:16}
  \frac{d}{dt} \zeta_1 
  &=
  \frac{\kappa_{p_1}}{\beta} 
  \left( 
    \frac{p_1}{m} G_{p_1} 
    \frac{e^{\beta\hbar\omega}-1}{\hbar\omega} -
    \frac{\partial G_{p_1}}{\partial p_1} +\varepsilon
    G_{p_1} \frac{p_1-p_2}{m \hbar\omega} \frac{1}{e^V +\varepsilon 1}
  \right)
  \\
  \frac{d}{dt} \zeta_2 
  &=
  \frac{\kappa_{p_2}}{\beta} 
  \left( 
    \frac{p_2}{m} G_{p_2} 
    \frac{e^{\beta\hbar\omega}-1}{\hbar\omega} -
    \frac{\partial G_{p_2}}{\partial p_2} -\varepsilon
    G_{p_2} \frac{p_1-p_2}{m \hbar\omega} \frac{1}{e^V +\varepsilon 1}
  \right)
  \nonumber
  \\
  \frac{d}{dt} \xi_1 
  &=
  \frac{\kappa_{r_1}}{\beta} 
  \left( 
    m \omega^2 r_1 F_{r_1} 
    \frac{e^{\beta\hbar\omega}-1}{\hbar\omega} -
    \frac{\partial F_{r_1}}{\partial r_1} +\varepsilon 
    F_{r_1} m \hbar\omega (r_1-r_2) \frac{1}{e^V +\varepsilon 1} 
  \right)
  \nonumber
  \\
  \frac{d}{dt} \xi_2 
  &=
  \frac{\kappa_{r_2}}{\beta} 
  \left( 
    m \omega^2 r_2 F_{r_2} 
    \frac{e^{\beta\hbar\omega}-1}{\hbar\omega} -
    \frac{\partial F_{r_2}}{\partial r_2} -\varepsilon
    F_{r_2} m \hbar\omega (r_1-r_2) \frac{1}{e^V +\varepsilon 1}
  \right)
  \ .
  \nonumber 
\end{align}
The effects of Pauli-blocking and Bose-attraction are now present in
the equations of motion of all pseudo\-friction coefficients.
%%%%%%%%%%%%%%%%%%%%%%%%%%%%%%%%%%%%%%%%%%%%%%%%%%%%%%%%%%%%%%%%%%%%%%%%
\section{Results}

\subsection{Bose-attraction and Pauli-blocking}
The different signs for bosons and fermions in the equations of motion
of the pseudofriction coefficients stem from the respective
two-particle wavefunction. We investigate the consequences for the
movements of the particles. Obviously, the effects of Bose-attraction
and Pauli-blocking will be most pronounced at low temperatures, when
both particles tend to occupy the one-particle ground state and
thereby get close to one another in phase space.
 
We examine the most elementary example, the scheme given by the set of
equations of motion \eqref{eq:7} and \eqref{eq:12}. When two
\emph{fermions} are close in phase space, $V=|\alpha_1-\alpha_2|^2$
becomes very small and the factor $1/(e^V-1)$ gets very large, causing
a strong acceleration of $\petaeins$ and $\petazwei$, and thereby of
$p_1$ and $p_2$ into opposite directions due to the different signs.
Effectively, a close approach of the particles in phase space,
corresponding to $V \to 0$, is avoided by the dynamics. Thus, in the
case of fermions, the additional terms in the equations of motion
\eqref{eq:12} act like a repulsive force.

\begin{figure}
  \begin{center}
    \epsfig{file=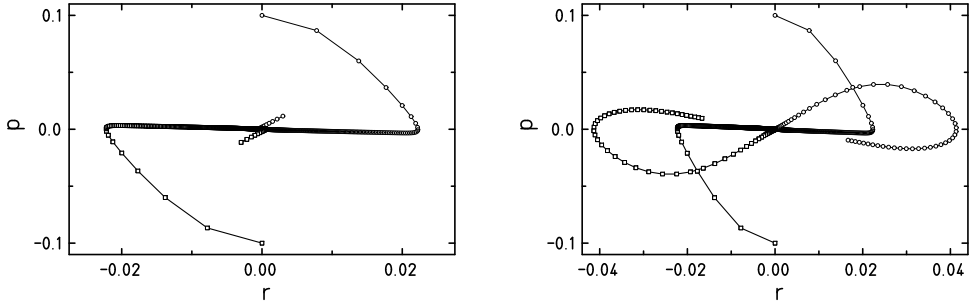,width=130mm}
    \caption{Left panel: isothermal dynamics of two bosons,
      right panel: two fermions. The initial values of $ \{
      r_1,p_1,r_2,p_2 \} $ are $ \{ 0,-0.1,0,0.1 \} $ in both cases,
      the temperature is $T=0.1 \hbar\omega$, and $Q_1=Q_2=0.5$. The
      integration time is $8.1$ periods of the harmonic oscillator for
      the bosons, and $6.5$ for the fermions. The time distance
      between the symbols is $0.013$. The figure illustrates the
      effect of the Bose attraction and the Pauli-blocking in the
      dynamics: While the bosons stay close to each other for a longer
      time, the fermions are immediately driven away from each other
      due to the exclusion principle.}
    \label{fig:1}
  \end{center}
\end{figure}

In the case of bosons, the opposite signs in equations \eqref{eq:7},
\eqref{eq:12} cause an acceleration of the parameters in the direction
of one another, favoring a ``meeting'' of the particles in phase
space. So the case $V=0$ is not excluded at all; on the contrary, it
is aimed at with the maximum value of $1/(e^V+1)$. Fig.~\ref{fig:1}
illustrates the typical behaviour of two identical particles
neighboring in phase space.

In essence, the resulting effect of indistinguishability on the
thermostatted dynamics of identical quantum particles looks like an
attractive or repulsive interaction, although we treat a system of
non-interacting particles. The interaction which is of purely
statistical origin is mediated by the influence of the
pseudofrictional forces. Nevertheless, one can hope that this
statistical interaction influences the ergodicity of the system. As an
extreme example, one could think of thermalizing only one particle and
hope that the second one thermalizes due to the interaction, see next
section.

\subsection{Ergodicity}
The classical Nos\'e-Hoover method applied to a single particle in a
harmonic oscillator features ergodicity problems that are encountered
in the quantum case, too, since the overall characteristics of the
equations of motion are similar in both cases. However, there is no
classical counterpart of the quantum statistical interaction between
identical particles. Therefore, the question of whether the more
complex equations of motion presented in the precedent section lead to
improved ergodic behavior deserves attention.

For the different methods, we look at marginal distributions of the
thermal weight function $f_\varepsilon^{(2)}$. As an example, we give
the analytical expression of the marginal distribution of $r_1$:
\begin{align}
  \label{eq:17}
  f_\varepsilon(r_1)&=
  \frac{1}{\tilde{Z}_\varepsilon^{(2)}} 
  \int \frac{\dint p_1}{2 \pi \hbar} \, 
  \frac{\dint r_2 \, \dint p_2}{2 \pi \hbar} \;
  w^{(2)}_\varepsilon(r_1,r_2,p_1,p_2) 
  \\
  &=
  \frac{1}{\tilde{Z}_\varepsilon^{(2)}}
  \sqrt{\frac{m\omega}{8 \pi \hbar}} \Bigg( 
  \sqrt{\frac{1}{(e^{\beta\hbar\omega}-1)^3}} \;
  e^{-\frac{m\omega}{2\hbar} (e^{\beta\hbar\omega}-1) r_1^2} 
  \nonumber
  \\
  & \phantom{=
    \frac{1}{Z_\varepsilon^{(2)}}
    \sqrt{\frac{2 \pi m\omega}{\hbar}} \Bigg( }
  + \varepsilon
  \sqrt{\frac{1}{e^{2\beta\hbar\omega}-1}}
  \sqrt{\frac{1}{e^{\beta\hbar\omega}}}
  e^{-\frac{m\omega}{\hbar}\sinh(\beta\hbar\omega) r_1^2} \Bigg)
  \nonumber
  \ .
\end{align}

\subsubsection{Nos\'e-Hoover and Nos\'e-Hoover chain method}

\begin{figure}
  \begin{center}
    \epsfig{file=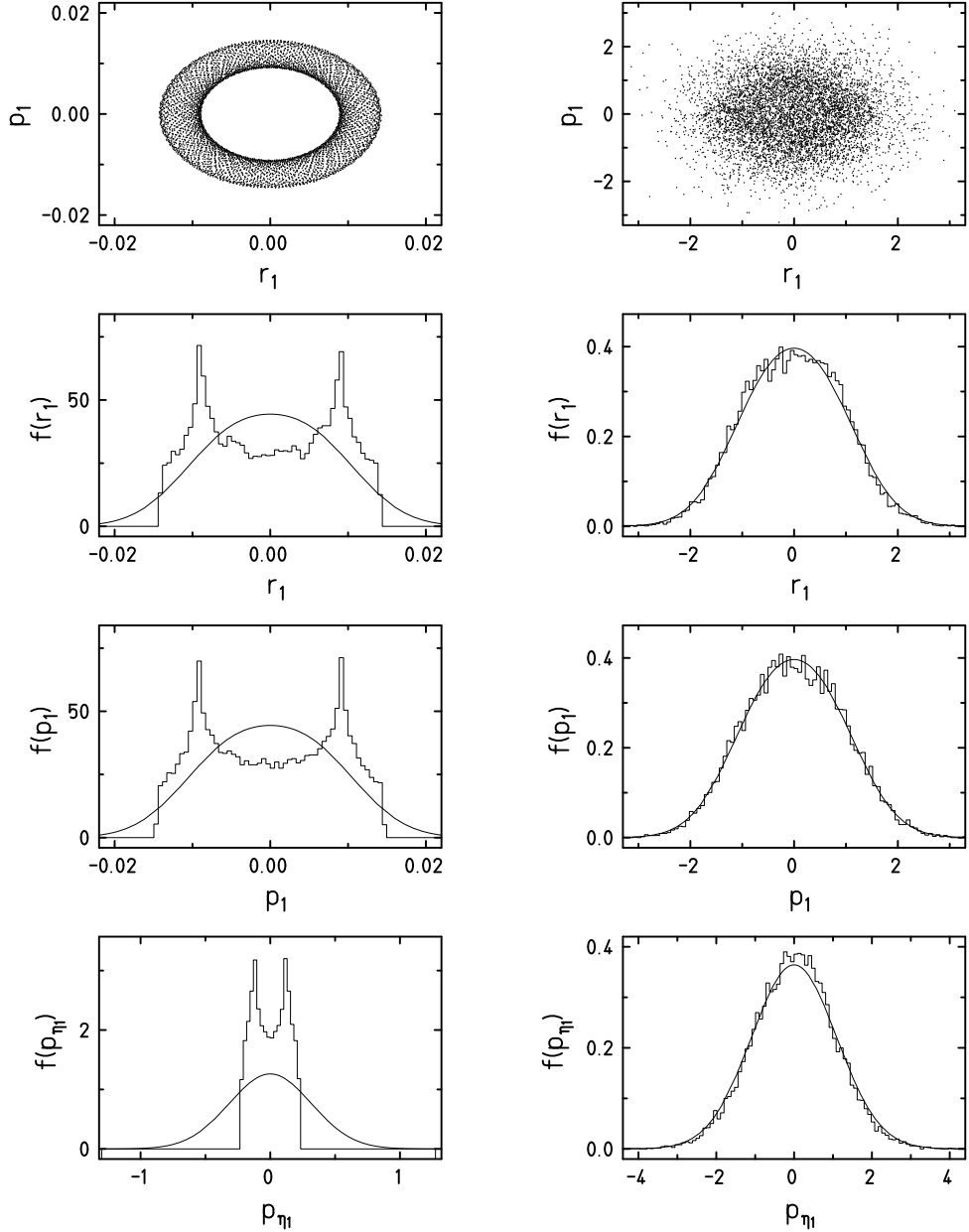,width=130mm}
    \caption{Results of time averaging with the simple
      Nos\'e-Hoover scheme for two fermions, left panel: $T=0.1$,
      right panel: $T=1.2$.
      In both cases, we used identical initial
      conditions, $r_1(0)=-r_2(0)=0.01$, $p_1(0)=-p_2(0)=0.01$,
      $\petaeins(0)=-\petazwei(0)=0.01$, $Q_1=Q_2=1.0$. We present
      density plots and histograms for $r_1$, $p_1$, and $\petaeins$;
      the respective results for $r_2$, $p_2$, and $\petazwei$ look
      similar.}
    \label{fig:2}
  \end{center}
\end{figure}

Firstly, we present results that are obtained using the equations of
motion \eqref{eq:7}, \eqref{eq:12}, i.~e.~the original Nos\'e-Hoover
technique generalized to the case of two identical fermions. A
detailed investigation of these equations of motion for a wide
temperature range shows that the system is not ergodic in general. We
find non-ergodic motion at low and intermediate temperatures $T
\lesssim 1.2$. However, above that value, we find ergodic motion, and
the respective marginal distributions are well matched by the
histograms obtained by time averaging, see Fig.~\ref{fig:2}. This fact
is to be contrasted sharply to the findings in the single-particle
case, where the simple Nos\'e-Hoover scheme produces non-ergodic
motion even at very high temperature values \cite{Men01,Hoo85}.

The different behavior of the two-particle dynamics may be attributed
to the statistical interaction appearing in this case. The more
complicated form of the equations of motion improves the ergodicity of
the scheme substantially compared to the case of a single particle.

As a result, this technique turns out to be not recommendable in the
low-temperature regime. Therefore, we have investigated whether the
problems of non-ergodicity can be overcome by standard methods that
are known from classical molecular dynamics, e.~g.~with a chain of
thermostats \cite{Mar92}.

Using one additional thermostatting pseudofriction coefficient for
each $p_{\eta_i}$, $i=1,2$, the problems of ergodicity are immediately
and reliably resolved. The marginal distributions are exactly matched
by the respective histograms, and the system is ergodic within the
whole large temperature range that has been investigated \cite{PhDMentrup}.
As in the classical case, it is remarkable that a solution to the
seemingly inaccessible problem of ergodicity is so easily obtained.
Moreover, from the point of view of computational time, it is
relatively inexpensive, since only two more degrees of freedom are
added, which is negligible especially if one deals with larger
systems.

We note that it is not sufficient to couple a second thermostating
pseudofriction coefficient to only one parameter, say $\petaeins$. In
this case, the marginals of $\petazwei$ are not sampled correctly at
temperature values $T \lesssim 0.3$. Even worse, if only $p_1$ is
coupled to a thermostat, the dynamics of the parameters $r_2$ and
$p_2$ is not affected at all. Hence, the energy of particle $2$ is an
additional conserved quantity, leading to strongly non-ergodic motion.
This illustrates the limits of the improved ergodicity due to the
statistical interaction.

\subsubsection{KBB method with different coupling schemes}

\begin{figure}
  \begin{center}
    \epsfig{file=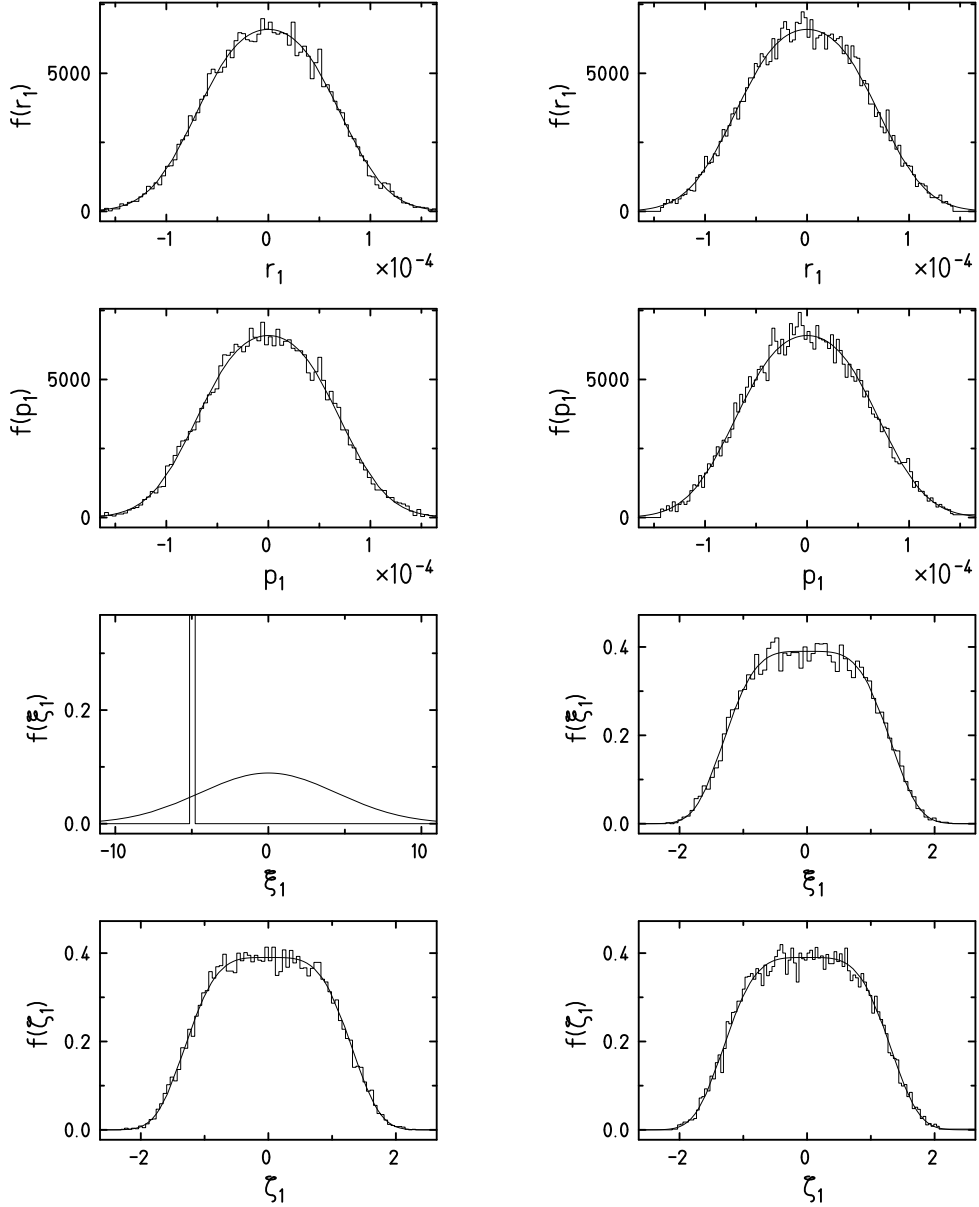,width=130mm}
    \caption{Marginal distributions obtained at $T=0.05$ with a cubic coupling
      scheme (left panel), and with the alternate scheme described in
      the text (right panel). The initial value of $\xi_1$ on the left
      panel is $\xi_1(0)=-5.1$, and the ``freezing'' of the demon
      variable is striking. Surprisingly, all other marginals are well
      reproduced. The right panel illustrates that the problem is
      resolved by the modified scheme.}
    \label{fig:3}
  \end{center}
\end{figure}

In the original paper \cite{KBB90}, Kusnezov, Bulgac, and Bauer
propose the following choice of functions
\begin{align}
  \label{eq:18}
  g_{r_1}(\xi_1)=\frac{1}{2}\xi_1^2 \ , \quad
  g_{r_2}(\xi_2)=\frac{1}{2}\xi_2^2 \ , \quad
  g_{p_1}(\zeta_1)=\frac{1}{4}\zeta_1^4 \ , \quad
  g_{p_2}(\zeta_2)=\frac{1}{4}\zeta_2^4 \ ,
\end{align}
as a reliable scheme that provides ergodic behavior. Using their rules
of thumb for the adaptation of the respective coupling constants, we
find that this cubic coupling scheme leads to ergodic motion also in
the present case. However, considering the low-temperature results for
fermions presented in Fig.~\ref{fig:3}, it can be inferred that the
marginal distributions of the pseudofriction coefficients $\xi_1$, and
equivalently, $\xi_2$ (not shown), that are linearly\footnote{Recall
  that the \emph{derivatives} of the functions given in \eqref{eq:18}
  appear in the equations of motion.} coupled to the system are not
sampled properly in the range $T\lesssim 0.1$. The histograms
presented in Fig.~\ref{fig:3} clearly indicate that $\xi_1$ does not
depart substantially from its initial value during the time evolution.
Nevertheless, the thermal weight function $w_\varepsilon^{(2)}$ is
sampled correctly, since all marginal distributions coincide with the
exact result. However, it is obvious that the cubic coupling scheme is
not ergodic in a part of the phase space at low temperatures. This
applies to bosons as well.

We propose to resolve this problem by coupling the coefficents
$\xi_1$ and $\xi_2$ in precisely the same manner as $\zeta_1$
and $\zeta_2$, i.~e.~cubically. This appears to be a sensible
improvement since the distributions of $\zeta_1$ and $\zeta_2$
are sampled correctly even at very low temperatures. The
resulting equations of motion yield the outcome shown in the
right hand panel of Fig.~\ref{fig:3}. All marginal distributions
are now sampled correctly.

\subsection{Two-particle density}
\label{sec:two-particle-density}

\begin{figure}[hhhhh!!!] 
  \begin{center} 
    \includegraphics[width=130mm]{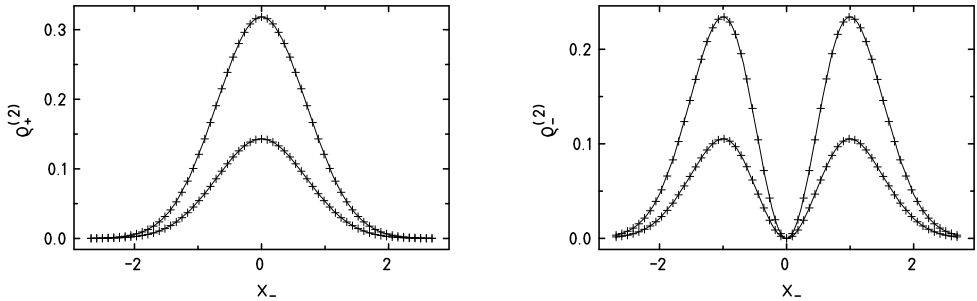}
    \caption{Results of time averages for the bosonic (left figure) and
      fermionic (right figure) two-particle density. The solid curves
      correspond to the respective analytical results as given by
      equation \eqref{eq:20}. In the fermionic case, the crosses are
      obtained from the simulations that gave the right hand panels of
      figure \ref{fig:3}. The upper line corresponds to the value
      $x_+=0$, the lower line to the value $x_+=0.894$.}
    \label{fig:4}
  \end{center}  
\end{figure}

As an example for the determination of the canonical average of a
two-particle observable, we present results of time averaging for the
two-particle density,
\begin{align}
  \label{eq:19}
  \rho_\varepsilon^{(2)}(x_1,x_2) = 
  \Mean{
    \frac{|\braket{x_1,x_2}{A_\varepsilon}|^2}
    {\braket{A_\varepsilon}{A_\varepsilon}}}
  \ ,
\end{align}
which can also be calculated analytically. In terms of the relative
and center-of-mass variables $x_-=\frac{1}{\sqrt{2}}(x_1-x_2)$,
$x_+=\frac{1}{\sqrt{2}}(x_1+x_2)$, the result is given by
\begin{multline}
  \label{eq:20}
  \rho_\varepsilon^{(2)}(x_-,x_+)
  =
  \frac{1}{\tilde{Z}_\varepsilon^{(2)}} 
  \frac{m\omega}{2\pi\hbar}
  \frac{1}{e^{2\beta\hbar\omega}-1}
  \exp \left(
    - \frac{m \omega}{\hbar} \tanh \Big(\half \beta \hbar \omega \Big) 
        (x_+^2+x_-^2)
  \right) \\
  \times\; \left( 
    1 + \varepsilon
    \exp\Big(
      -4 \frac{m\omega}{\hbar} 
      \frac{e^{\beta\hbar\omega}}{e^{2\beta\hbar\omega}-1} 
      \, x_-^2
    \Big)
  \right)
  \ .
\end{multline}
In order to give a useful representation of our findings, we present
results of time avera\-ges along with the respective analytical
results for fixed values of $x_+$ that have been chosen arbitrarily.
Fig.~\ref{fig:4} shows an excellent agreement.
%%%%%%%%%%%%%%%%%%%%%%%%%%%%%%%%%%%%%%%%%%%%%%%%%%%%%%%%%%%%%%%%%%%%%%%%
\section{Summary and outlook} 

This article presents an extension of the powerful methods of heat
bath coup\-ling in classical MD simulations to a quantum system of
identical particles. Surprisingly, the classical ansatz turns out to
be suitable for the sampling of quantum entangled distribution
functions. The resulting two-particle dynamics contains additional
terms that act like an attractive (bosons) or repulsive (fermions)
force mediated by the pseudofriction coefficients.  Ergodicity
problems are reliably resolved by Nos\'e-Hoover chains or a
modification of the cubic coupling scheme, even at very low
temperature values. An analytical treatment of the case of an
$N$-particle Fermi system is possible, and the resulting equations are
given in Ref.~\cite{PhDMentrup}.

The most promising prospect that our method offers lies in a
combination with approximate quantum dynamics schemes. A variety
of such schemes is available, some of which are based on the
time-dependent quantum variational principle \cite{KrS81} which
allows to derive approximations to the time-dependent
Schr\"odinger equation. How can we combine the thermostating
method developed in this work with, e.~g., Fermionic Molecular
Dynamics (FMD) \cite{RMP} in order to obtain an isothermal
dynamical scheme for a complex interacting fermion system? Does
our generalization of the classical methods to quantum dynamics
permit to make the power of approximate quantum MD schemes
available for the calculation of quantum canonical averages
without diagonalising the full many-body Hamiltonian? This
question is very timely in view of recent experiments
investigating the behavior of trapped Fermi gases.

In FMD, the variational trial state is a Slater determinant of
single-particle Gaussian wave packets parametrized by mean position,
mean momentum, and complex width. These wave packets are frequently
referred to as {\it squeezed states} and may be regarded as
generalisations of coherent states. Therefore, an application of the
thermostats developed in the present work to FMD appears feasible.

Another idea of combining FMD with a thermostat is to cool the system
of interest ``sympathetically'', i.~e.~via an interaction between
particles that are kept at a constant temperature by a quantum
Nos\'e-Hoover chain and the physical system under investigation. This
corresponds precisely to the experimental technique of ``sympathetic
cooling'' currently employed to investigate ultracold fermionic gases
\cite{Myatt97,DeMarco98}. Since the thermalizing of the particles
coupled to a Nos\'e-Hoover chain would correspond to a thermalizing of
non-interacting particles, such a method is only approximately correct
since we need to employ an interaction to enable the sympathetic
cooling.

Despite these difficulties, an important asset of an isothermal MD
scheme is that it can possibly provide temporal information, in
particular time correlation functions. Although it is not clear to
which degree the extended system methods realistically mimic the heat
bath interaction, the underlying equations of motion are physically
reasonable. Therefore, in principle, this method is tailor-made to
model the particle dynamics at constant temperature in a magnetic
trap. 
%%%%%%%%%%%%%%%%%%%%%%%%%%%%%%%%%%%%%%%%%%%%%%%%%%%%%%%%%%%%%%%%%%%%%%%%

{\bf Acknowledgments}\\[5mm] The authors would like to thank the
Deutsche Forschungsgemeinschaft (DFG) for financial support of the
project ``Isothermal dynamics of small quantum systems".
%%%%%%%%%%%%%%%%%%%%%%%%%%%%%%%%%%%%%%%%%%%%%%%%%%%%%%%%%%%%%%%%%%%%%%%%

\end{document}